\documentclass[12pt,preprint]{aastex}
\usepackage{emulateapj5}
\usepackage{apjfonts}

\submitted{Accepted for publication in the Astrophysical Journal}

\newcommand{\swin}{Centre for Astrophysics and Supercomputing, Swinburne
University of Technology, P.O.~Box 218 Hawthorn, VIC 3122, Australia}

\newcommand{\etal}{et al.\ }

\begin{document}
\title{A Search for Sub-Millisecond Pulsars}
\author{R. T. Edwards, W. van Straten and M. Bailes}
\affil{\swin}
\email{redwards@astro.uva.nl}
\begin{abstract}
We have conducted a search of 19 southern Galactic globular clusters
for sub-millisecond pulsars at 660~MHz with the Parkes 64-m radio
telescope. To minimize dispersion smearing we used the CPSR baseband
recorder, which samples the 20~MHz observing band at the Nyquist
rate. By possessing a complete description of the signal we could
synthesize an optimal filterbank in software, and in the case of
globular clusters of known dispersion measure, much of the dispersion
could be removed using coherent techniques. This allowed for very high
time resolution (25.6~$\mu$s in most cases), making our searches in
general sensitive to sub-millisecond pulsars with flux densities
greater than about 3~mJy at 50~cm. No new pulsars were discovered,
placing important constraints on the proportion of pulsars with very
short spin periods in these clusters.
\end{abstract}


\section{Introduction}
With the discovery of the first `millisecond' pulsar, B1937+21
($P=1.56$~ms; \citealt{bkh+82}), it became clear that neutron stars
can achieve very rapid rotation rates. This period is close to but
still somewhat greater than the minimum rotation period for a neutron
star, below which the star becomes unstable to mass shedding at the
equator (see e.g. \citealt{cst94}). Various equations of state have
been proposed for nuclear matter and due to the differences in density
under such equations, the limiting spin period of neutron stars
depends on the choice of equation of state. The discovery of a pulsar
with $P < 1$~ms would be of great value in eliminating potential
equations of state.  The distribution of periods of known pulsars cuts
off quite sharply below about $P=2$~ms, however it is not clear
whether this cut-off is intrinsic to the pulsar population (see
\citealt{llb+96}). Most previous pulsar surveys had a time resolution
of $\sim 300$~$\mu$s for nearby pulsars, resulting in a strong decline
in sensitivity for periods less than a few milliseconds, indicating
that the short-period cut-off may in fact be a selection effect. The
only effective way to answer this question is to conduct surveys with
sufficient time resolution to be well-sensitive to pulsars with $P <
1$~ms, and preferably with a flat sensitivity response to periods well
below a millisecond in order to eliminate any bias towards longer
pulse periods. For a traditional analog filterbank system this would
require a large number of channels with fast sampling and would
involve considerable cost. An alternative is to record the raw
receiver voltages at baseband and to perform all frequency decimation
and detection in software for the best possible time and frequency
resolution. Until recently this approach has been little used due to
the formidable data storage and processing requirements, however the
rapid development that has occured in these areas in the last decade
means that the required hardware is now relatively affordable.  We
report here for the first time the use of baseband processing in a
pulsar survey.

Historically approximately half of all millisecond pulsar discoveries
were made in Globular clusters, with most of the remainders found in
large scale surveys of the Galaxy. Since the cores of most Globular
clusters are easily contained in a single telescope beam for low to
intermediate observing frequencies, globular cluster searches are very
efficient in their return of millisecond pulsars for observing
campaigns of limited duration. Due to the large amounts of data to be
processed from observations of high time and frequency resolution,
globular clusters are a logical place to begin the search for
sub-millisecond pulsars. We conducted several observations of each of
19 southern globular clusters with the CPSR baseband recording system
\citep{vbb00} at the Parkes radio telescope. This paper describes the
searches and their results.

\section{Observations and Analysis}
In an observing run of four days from 2000 March 17--20 we conducted
observations of 19 southern Galactic globular pulsars with the Parkes
64-m radio telescope. The signals from two orthogonal linear
polarizations of the 50-cm receiver were mixed to baseband in a
quadrature down-convertor and filtered to provide a 20~MHz band
centered at a sky frequency of 660~MHz.  The in-phase and quadrature
components in each polarization were 2-bit sampled at a sample rate of
20~Msamples~s$^{-1}$ in accordance with the Nyquist theorem for
complete description of the band-limited signal. Sampler thresholds
were set at the beginning of the observation in accordance with the
prescriptions of \citep{ja98} and were held at these values for the
duration of each observation.  The resultant data stream had a
bit-rate of 160 Mbit~s$^{-1}$ and was written in segments of $\sim53.7$~s
(corresponding to 1~GB of data) to DLT 7000 tapes. The recording
system consisted of four DLT 7000 drives, a large disk array and a Sun
Ultra 60 workstation and in conjunction with the down-conversion and
sampling systems is known as the Caltech-Parkes-Swinburne Recorder
(CPSR) (van Straten \etal, in preparation). A total of 55 observations
of 30 minutes' duration each were recorded, resulting in a 1.8~TB
data-set.

The observations were processed on the 64-node Swinburne workstation
cluster. Each observation could be processed by one of the schemes
described below in about 12 hours using 12 500~MHz Compaq EV6 processors.
Tapes were unloaded to a 1-TB RAID array to facilitate the
reassembly of full 30-minute contiguous data streams from the 1~GB
files that were distributed across four or more tapes during recording.
The statistics of samples in each file were examined and any files
partially recorded before the sampler levels were correctly set were
discarded.

In order to allow an efficient search in dispersion measure (DM), a
filterbank was synthesized in software. The data were loaded in
segments and each polarization in each segment was transformed to the
Fourier domain by means of an FFT.  The spectrum was then divided into
512 equal sections and each section was transformed back to the time
domain with an inverse FFT to produce a number of time samples in each
of 512 frequency channels. The squared magnitude of each sample was
taken and samples were summed in polarization pairs.  The resultant
sample interval in each channel was $25.6$~$\mu$s.

In the case of observations of clusters with previously known pulsars,
we convolved each channel with a filter matched to the inverse of the
response function of the interstellar dispersion at the estimated
dispersion measure to the cluster \citep{hr75} by multiplication while
the data was still in the Fourier domain.  In this case, the length of
the original FFT was important for it determined the effective
resolution of the de-dispersion filter to be applied. In addition, the
cyclical nature of the convolution with the filter dictated that a
certain number of samples (corresponding to approximately half the
dispersive delay across each channel) must be discarded from either
end of the resultant time series.  To recover these samples the
segments were overlapped, resulting in some inefficiency that could
be minimized through the use of larger segments
(transforms). Typically segments of $2^{15}$ samples were used,
with a 64-point coherent de-dispersion filter applied to each of the
512 channels.

The resultant data was similar in form to that recorded from orthodox
analog filterbanks, with the exception that each channel was recorded
at half its Nyquist rate\footnote{The full Nyquist rate could be
achieved through the use of complex-to-real inverse transforms,
however the time resolution was deemed sufficient as it stood.}, much
faster than commonly practiced with analog filters. In addition, the
use of coherent de-dispersion in clusters with known pulsars
meant that the dispersion smearing induced across each channel arose
only from the difference between the true and assumed dispersion
measure, and was generally expected to be much shorter than the sample
interval. The dispersion smearing experienced by pulsars in clusters
lacking a DM estimate is given approximately by DM$/22.7$~pc~cm$^{-3}$
in units of samples of 25.6~$\mu$s in duration, where
$22.7$~pc~cm$^{-3}$ is the so-called `diagonal' DM.

The synthesized filterbank data for each observation were partially
de-dispersed with the `tree' algorithm of \citet{tay74} before being
subject to a search process similar to that used for the intermediate
latitude multibeam survey \citep{ebvb01}. Data were de-dispersed and
summed into time series at a range of trial dispersion measures,
either about a nominal cluster DM where such a value was available, or
up to a maximum of 732 pc~cm$^{-3}$ in 1463 steps.
The spacing of these steps was such that the pulse smearing induced
in pulsars at DMs half-way between two trial values was comparable to
the smearing across a single filterbank channel.
It was possible that the motion of a pulsar in its
orbit about a close companion could have induced significant change in the
observed pulse period over the course of the observation, and for this
reason we incorporated an acceleration search. For each trial
dispersion measure we computed several time series re-binned in such a
way as to compensate for the effects of a range of values of
acceleration. The method used was similar to that described by
\citep{clf+00}. The resulting time series with trial pairs of values
for dispersion measure and acceleration were then searched for
periodicities and a page of diagnostic information was produced for
each potentially significant candidate.

The performance of much of the signal processing in software allows
for a considerable degree of flexibility. We used several different
sets of parameters as appropriate depending on whether the cluster had
a known pulsar and also on the available computer time. The scheme
denoted FD (`full DM search') searches the full range of dispersion
measures from 0 to 732 pc~cm$^{-3}$ in 1463 steps with integrations
of 1718~s. For clusters with
known DM, the scheme known as FA (''full acceleration search'')
applied the coherent de-dispersion kernels and searched a small
specified range of dispersion measures in 201 accelerations evenly
spaced in the range given by $|a| \leq 30$~m~s$^{-2}$. Due to the
large computational burden of the task, the FA search only processed
859~s of data, using $2^{25}$-point transforms. Finally, to maintain
sensitivity to accelerated millisecond pulsars in clusters with no DM
estimate, we also performed a search denoted SA (`slow acceleration
search') which synthesized a filterbank of 256 channels and summed
the resultant detected samples in groups of sixteen.  The resultant
sample rate was $204.8$~$\mu$s with a diagonal DM of
$90.8$~pc~cm$^{-3}$, and the full range of 0 to 2950~pc~cm$^{-3}$ was
searched at 737 trial values.  Again, only 859~s of data were
processed, with $2^{22}$-point transforms and 121 accelerations with
$|a| \leq 30$~m~s$^{-2}$.

A total of 10 clusters with previously known pulsars and 9 without
were searched as indicated in Tables 1 and 2. The latter were selected
from the catalog of \citet{har96} on the basis of proximity and
luminosity. Since \citet{fg00} observed several globular clusters in
radio continuum and found unidentified steep spectrum emission in NGC
6544, and Liller 1 it was ensured that these clusters were included in
the selection.  The general strategy was to process clusters with an
accurate DM estimate using the FA parameters, and those without using
the SA parameters. For the FA searches the dispersion measure range
was generally selected in such a way as to include previously
published values for pulsars in the cluster and to allow for $\pm
1$~pc~cm$^{-3}$ of variation due to gas in the cluster environment.
Since the published dispersion measure for PSR B1745--20 in NGC 6440
is significantly uncertain \citep{mlj+89b}, data from this cluster were
processed with the SA parameters, as were those clusters lacking a
previously known pulsar.  Observations of all clusters were also
processed with the FD parameters. For clusters also processed using
FA, the FD search offers sensitivity outside the limited DM range of the
FA search, whilst for those processed with SA the additional FD
processing provides sensitivity to sub-millisecond pulsars. However,
in both cases the FD search is insensitive to significantly
accelerated pulsars.  

Due to the frequent loss of data stored on DLT
tapes, some observations were not processed to the full extent of the
scheme above. In particular, only three of the four observations of
NGC 4833 were processed in SA mode, and only four of the five
observations of NGC 6624 and neither of the two observations of NGC
6752 were processed in FD mode. At the start of processing, no pulsars
were known in NGC 6752, so it was searched using the SA
parameters. With the announcement of the discovery of a pulsar in this
cluster \citep{dlm+00b} it was decided to re-process the data using the FA
parameters and the published dispersion measure, however only one
observation was successfully retreived from tape and processed in this
manner.

\section{Results and Discussion}
\subsection{Detections}
Offline processing yielded detections of six pulsars, all of which
were previously known, including PSR B1744--24A in Terzan 5 with a
line-of-sight acceleration of 29~m~s$^{-2}$. The pulsars are listed in
Table \ref{tab:det} along with their pulse periods and the signal to
noise ratio of detections. Variations in flux density due to
scintillation (and in the case of PSR B1744-24A, eclipses) are the
expected cause of variations in detectability and signal-to-noise
ratio from one observation to the next.  

One promising pulsar candidate was observed in 47 Tucanae in an
observation centered at MJD 51621.25740, with a topocentric period of
3.756394~ms, a dispersion measure of 24.3~pc~cm$^{-3}$ and an
acceleration of 7.2~m~s$^{-2}$. The signal-to-noise ratio of this
candidate was 9.7, placing it at the threshold of credibility, and in
the absence of any other detections of similar periodicities we are
hesitant to label the signal a `pulsar'. All other periodicities were
consistent with random chance (due to the complexity of the search
space) or persistent terrestrial interference.



\subsection{Sensitivity}
The sensitivity of pulsar observations is a function of the radiometer
noise and the observed duty cycle. After \citet{dtws85}, the minimum
detectable mean flux density is

\begin{equation}
S_{\rm min} = 
	\frac{\alpha\beta T_{\rm sys}}
	     {G\sqrt{N_{\rm pol}Bt_{\rm obs}}}
	\sqrt{\frac{\delta}{1-\delta}}
\end{equation}
where $\alpha$ is a dimensionless loss factor, $\beta$ is the
threshold signal to noise ratio, $T_{\rm sys}$ is the system
temperature, $G$ is the telescope gain, $N_{\rm pol}$ is the number of
polarizations, $B$ is the observing bandwidth, $t$ is the integration
time and $\delta$ is the effective duty cycle.  The observed pulses
are broadened somewhat relative to those actually emitted by the
pulsar due to factors such as multi-path propagation
(scatter-broadening), dispersion smearing in filterbank channels, and
the finite duration of the sampling interval. These effects should be
added in quadrature with the intrinsic pulse width to yield the
effective pulse width.  All pulsar systems incorporate a degree of
sensitivity loss modeled with $\alpha$, the major factor in the past
being the use of 1-bit sampling which contributes
$\sqrt{\pi/2}\simeq1.25$ to this value. Most surveys in the past have
assumed an extra 15 per cent loss due to other factors, giving
$\alpha=1.5$. Since the present survey samples with two bits of
precision, we begin with a value of $1.15$ (see \citealt{ja98}) and
adding 15 per cent for other losses, arrive at an assumed value of
$\alpha=1.3$. From the appearance of spurious signals in this search
and based on the distribution of true pulsar signal to noise ratios in
previous work \citep{ebvb01}, we use a value of $\beta=10$ as the
minimum signal to noise ratio for a pulsar candidate of firm
credibility. The effective system temperature is mainly the result of
contributions from thermal noise in the receiver and from Galactic
synchrotron radiation. The receiver used in this work contributes
approximately 60~K to the system temperature, whilst the Galactic
contribution is a strong function of Galactic latitude, contributing
$\sim$10~K at high Galactic latitudes and $\sim$100~K at the Galactic
plane.  Pulsar signals are superimposed on this noise with a gain from
the 64-m collecting dish of $G\simeq0.6$~K~Jy$^{-1}$.

The sensitivity as a function of period for the searches described
here is plotted in Figures \ref{fig:sensitivityF} and
\ref{fig:sensitivityS}. The values are for a cold sky of around 10~K;
for clusters near the Galactic plane sensitivity drops by around a
factor of 2.3.  Our analysis includes the effects of the finite sample
interval and dispersion smearing in filter channels, but does not
attempt to model scatter-broadening due to its dependence on the
composition of the intervening interstellar medium. From the work of
\citep{rmdm+97} we expect most pulsars in the clusters searched to
experience of the order of one sample of scattering (to $50\%$
intensity, based on a DM of 80 pc~cm$^{-3}$), however the spread
about this value is likely to be large.
It should also be noted
that sensitivity is strongly dependent on the intrinsic pulse width, a
factor which has been somewhat neglected in the past. Since the
majority of recycled pulsars have pulse duty cycles between 5 and 30
per cent (see e.g. \citealt{kxl+98}), we show in the figures sample
curves for these values, resulting in a baseline sensitivity of 1.3--4
mJy.  Figure \ref{fig:sensitivityF} applies to observations of 859~s
in duration with 512 channels and 25.6~$\mu$s sampling, such as the FA
search.  The curves for FD would be of the same shape but with a
slight downward shift (by a factor of $1/\sqrt{2}$ or $\sim$0.15
decades). The maximum DM range searched in the FA search was only
$\pm2.5$~pc~cm$^{-3}$ and the dispersion smearing in each channel even
at the edges of the range is very small due to the earlier coherent
removal of a nominal cluster DM. Hence the zero-DM curves in Figure
\ref{fig:sensitivityF} are the most appropriate for clusters with
known pulsars. For the SA search the curves in Figure
\ref{fig:sensitivityS} should be used. Note that the sampling interval
of this search configuration (204.8~$\mu$s) is comparable to previous
searches, and the degradation of sensitivity at high dispersion
measures is improved due to the small channel bandwidth (and hence the
large diagonal DM of $90.8$~pc~cm$^{-3}$). Nevertheless, it is
apparent from Figure \ref{fig:sensitivityS} that the available
sensitivity declines rapidly as periods go below 10~ms. On comparison
of Figures \ref{fig:sensitivityF} and \ref{fig:sensitivityS} the
superiority of high time resolution processing over traditional survey
configurations (analogous to SA) for the detection of millisecond and
sub-millisecond pulsars is clear.

However, it must be noted that whilst the present work had good
sensitivity to sub-millisecond pulsars, its sensitivity to slower
pulsars was comparable to or poorer than previous surveys.  The four
published 50-cm flux densities we have found in the literature for
previously known pulsars in these clusters are presented in Table
1. These flux values are compatible with our detections and
non-detections. PSR B1718--19 (in NGC 6342) lies well below the flux
limit and was not detected. The (rather uncertain) reported value for
PSR B0021--72C in 47 Tucanae, also un-detected, lies slightly above to
the predicted minimum given its duty cycle of $\sim$15\%, however we
expect that the reported deep scintillation of pulsars in 47 Tucanae
is the cause of this result as well as the strong detection of
B0021--72D, reported to be weaker than B0021--72C \citep{mlr+91}. The
remaining undetected pulsars were mainly discovered in work that was
either of greater intrinsic sensitivity, or that exploited
scintillation-induced flux variability by conducting numerous
observations of the clusters at different epochs.

\subsection{Acceleration Effects}
The preceding analysis neglects the effects of acceleration due to
orbital motion. The differential Doppler shift induced over the course
of the observation can result in the smearing of power across several
bins of the fluctuation spectrum. Since the shift moves frequency
components a by multiplicative factor ($at/c$), the loss of
sensitivity is greatest at higher harmonics. The acceleration spacing
of the FA search is $\sim$0.3~m~s$^{-2}$, implying a worst-case
differential shift factor of $0.15$~m~s$^{-2}\times 859$~s$/c \simeq
4.30\times 10^{-7}$ for a pulsar with an acceleration mid-way between
two trial values. Since each bin of the fluctuation spectrum
represents $1/859$~s~$\simeq 1.16\times 10^{-3}$~Hz, the 2500-Hz
fundamental of a putative $0.4$-ms sub-millisecond pulsar would
experience at most $\sim$1 bin of acceleration smearing, with higher
harmonics experiencing proportionately more smearing but representing
a smaller fraction of the total power of the pulsar. 

\begin{figure}
\epsscale{0.5}
\plotone{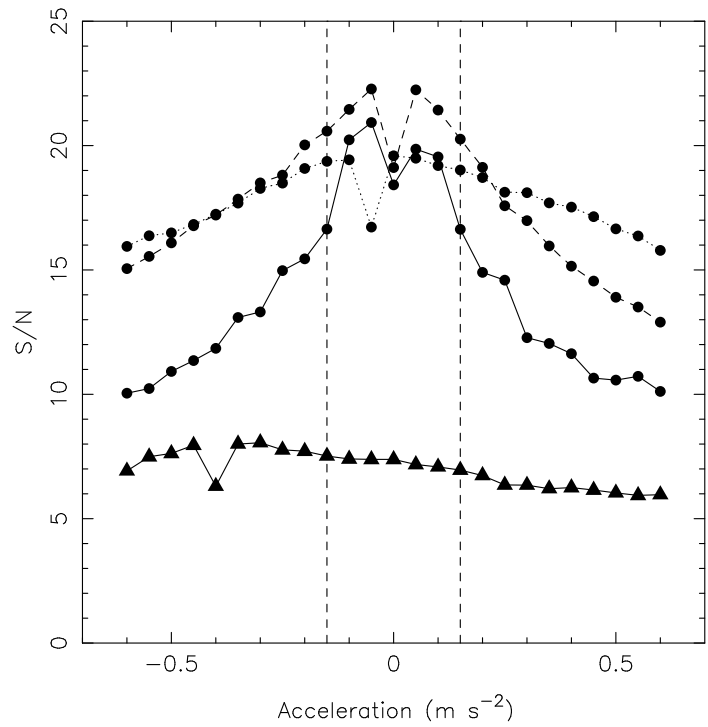}
\caption{Spectral signal signal to noise ratio as a function of trial
acceleration for simulated pulsar signals. Circles and triangles
represent values resulting from Gaussian profiles with FWHM values of
5 and 30 per cent respectively, and are joined by solid, dashed and
dotted lines for pulse periods of $\sim$0.43, 0.89 and
1.84~ms. All signals were simulated with zero acceleration and equal
mean flux density. Vertical dashed lines are placed at
$\pm$0.15~m~s$^{-2}$ to represent displacements from zero by half the
spacing used in the FA search.  }
\label{fig:accsn}
\end{figure}

To examine the true sensitivity loss we processed real data from an FA
search, to which we had added simulated signals from non-accelerated
pulsars of various pulse widths and periods. We used a trial
acceleration spacing of 0.05~m~s$^{-2}$ in the range
$|a|<6$~m~s$^{-2}$ to examine the loss of signal to noise ratio when
trial accelerations do not match the true acceleration of the pulse
(in this case zero). The results are shown in Figure
\ref{fig:accsn}. Even in the worst case, a 0.4~ms signal with a pulse
width of 5 per cent FWHM (that is, at the extreme lower end of
observed MSP pulse widths), some 80 per cent of sensitivity was
retained at the outer edges of the range of acceleration offsets
experienced in the FA search. It is clear that the impact of
acceleration on sensitivity of the FA search was small for $P \gtrsim
0.4$~ms.  In addition, due to scintillation it is expected that a real
pulsar would experience less signal-to-noise loss than is indicated by
these results. For the SA search a more modest acceleration spacing of
0.5~m~s$^{-2}$ was chosen due to the fact that the long sample
interval employed itself severely limits sensitivity to sub-millisecond
pulsars. In this case, any pulsar or harmonic with a period greater
than 0.6~ms would experience less than one bin of acceleration-induced
spectral smearing.

Of course, all of the above only applies to pulsars with accelerations
that remain essentially constant throughout the observation and lie
within the range of $|a|<30$~m~s$^{-2}$ searched. The former
requirement is likely to be satisfied for the vast majority of
systems; the loss of signal-to-noise ratio under a constant
acceleration approximation is likely to be significant only for
systems with orbital periods less than a few hours (see
\citealt{jk91}). Nevertheless, such exotic and interesting systems do
exist and it is important that general surveys maintain sensitivity to
them, either through repeated observation (to exploit favorable
orbital phases of nearly constant acceleration) or by the use of more
involved techniques (e.g. \citealt{ran00}).  

Of those systems
exhibiting constant acceleration, sensitivity in this work may still
have been compromised if the acceleration exceeded $\pm 30$m~s$^{-2}$.
This range is typical of previous pulsar searches and is likely to
encompass the majority of pulsar systems, however there are several
known exceptions.  The eclipsing binary of Terzan 5 \citep{lmd+90}
experiences line-of-sight accelerations greater than this for
approximately 30 per cent of its orbit, with a maximum value of
33.2~m~s$^{-2}$.  The eccentric double neutron star systems B2127+11C
(in M15), B1534+12, B1913+16 also experience strong accelerations,
exceeding 30~m~s$^{-2}$ in the line of sight for 20--50 per cent of
the orbit. Such systems have relatively long pulse periods ($P\gtrsim
30$~ms) and so would be detected in the present work at accelerations
up to around 40~m~s$^{-2}$. However the acceleration in these systems
exceeds even this value for a significant proportion of the time,
reaching a peak in excess of 100~m~s$^{-2}$ for the highly eccentric
systems B2127+11C and B1913+16. Eccentric systems with much shorter
orbital periods are expected to evolve from
pulsars typical of the presently known double neutron star population
via the loss of orbital energy in the form of gravitational radiation.
The detectability of such pulsars in this and previous globular
cluster searches would be severely affected by acceleration smearing.
The primary aim of this work was to detect or place limits on the
existence of [sub-]millisecond pulsars (which are not expected to have
neutron-star companions), and the computational load associated with
high resolution baseband processing limited the feasibility of
searching a broad acceleration range. However, we note that future
surveys with good basic sensitivity would be well-served by searching
a range of at least $\pm 100$~m~s$^{-2}$, perhaps at a reduced sample
rate of $\sim$1~ms and with correspondingly fewer trial dispersion
measures and accelerations.

\subsection{The Population of Sub-Millisecond Pulsars}
\begin{figure}
\epsscale{0.5}
\plotone{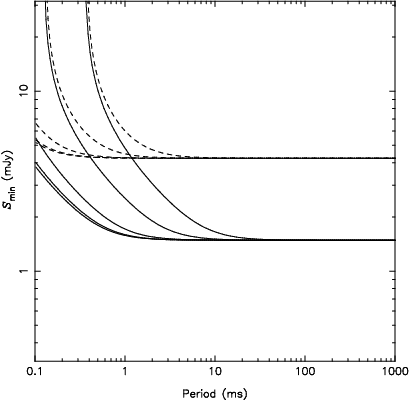}
\caption{Minimum detectable mean flux density as a function of pulse
period for FA search parameters. Solid and dashed
lines represent pulsars with 5 and 30 per cent intrinsic
duty cycles respectively. For each pulse width curves for dispersion
measures of 0, 10, 30, 100 and 300~pc~cm$^{-3}$ are shown
(from left to right).}
\label{fig:sensitivityF}
\end{figure}

\begin{figure}
\epsscale{0.5}
\plotone{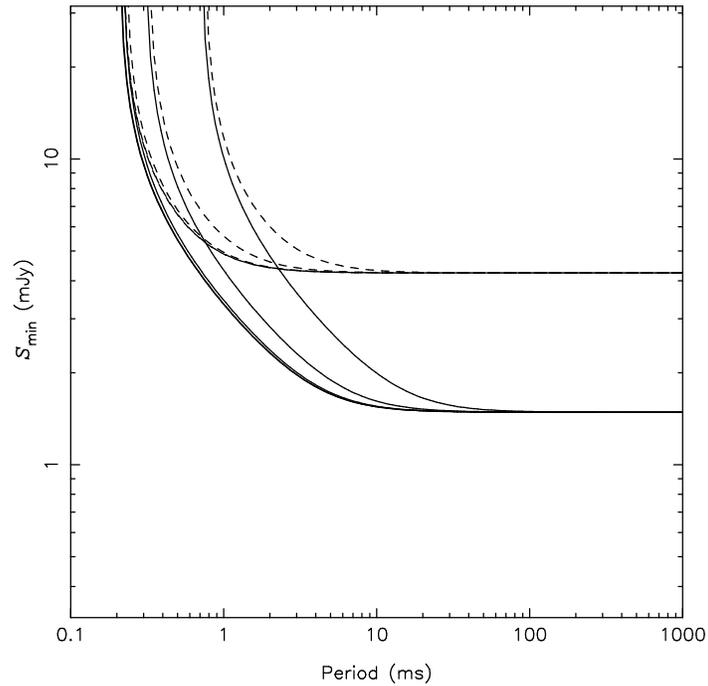}
\caption{Minimum detectable mean flux density as a function of pulse
period for SA search parameters. Solid and dashed lines represent
pulsars with 5 and 30 per cent intrinsic
duty cycles respectively. For each pulse width curves for dispersion measures of
0, 10, 30, 100 and 300~pc~cm$^{-3}$ are shown (from left to right),
although the first three are in general so close as to be difficult to
distinguish (c.f. Figure \ref{fig:sensitivityF}).}
\label{fig:sensitivityS}
\end{figure}

It is clear from inspection of Figure \ref{fig:sensitivityF} and the
preceding analysis that our FA search, unlike most searches in the
past, had a relatively flat sensitivity function for all periods
greater than $\sim0.4$~ms. 
The FD search also provided similar
characteristics for nearby (DM $\la 50$~pc~cm$^{-3}$) unaccelerated
pulsars. We are therefore in a stable position to analyze the period
distribution of the detected population relatively free of concerns
regarding selection effects. Unfortunately the system used was only
sensitive enough to detect a few pulsars, making any assertions
somewhat perilous due to small number statistics. However it is clear
even from this sample that the majority of recycled pulsars do have
pulse periods of a few milliseconds or more. All six pulsars detected
lay in the two octaves from 3--12~ms, whilst no pulsars were detected
in the preceding three octave interval, 0.375--3~ms over which we had
comparable sensitivity.

No new pulsars were discovered in NGC 6544, Liller 1, or Terzan 5
despite the presence of steep-spectrum emission as reported by
\citet{fg00}. It is probable that several pulsars in each cluster are
responsible for the emission, with each individual pulsar having a
flux density below our detection limit. An exception might be the `N'
source of Terzan 5 \citep{fg00}, which was unresolved at a resolution of
2\farcs9 (c.f. the cluster core radius of 11\arcsec;
\citealt{har96}). From the published spectral index and 20 cm flux, we
infer a 50 cm flux density of 9 mJy.  The sky temperature in this
region is $\sim 300$~K at 70 cm \citep{hssw82}, implying a temperature
of $\sim 100$~K at 50 cm \citep{lmo+87}. The sensitivity limits are
thus $160/70 \simeq 2.3$ times greater than indicated in Figure
\ref{fig:sensitivityF}, however even so, a very broad profile and/or
very short spin period would be required for the pulsar to have
remained undetected at 9~mJy.  It is possible that the source is a
pulsar and was undetected due to scatter broadening. Up to a
millisecond of scattering could be induced by the intervening
interstellar medium without having hampered the detection of the
presently known pulsars, with pulse periods of 11.6 and 8.4~ms
\citep{lmbm00}.  Such scattering would be catastrophic for the
detection of very fast pulsars and could explain the lack of detection
of the `N' source in this work.  Observations made at higher radio
frequency could answer this question since the time scale of scatter
broadening scales as $\nu^{-4}$. However, the lower flux density of
pulsars at high frequency necessitates the use of a large bandwidth,
beyond the present capabilities of baseband recording and processing
systems. Analog filterbanks generally do not have sufficiently narrow
channels to detect very fast pulsars at high dispersion measures such
as that of Terzan 5. Indeed, the probable non-detection at 20 cm of
the `N' source as a pulsar by \citet{lmbm00} must be taken with the
caveat that the system used induced $\sim$200~$\mu$s of dispersion
smearing, rendering it relatively insensitive to very fast pulsars.
Another alternative is that the `N' source is in a binary so close
(or with such a massive companion) that its line-of-sight acceleration
often exceeds the bounds of our search, or that its velocity evolution
is significantly non-linear on the timescale of our observations.

To strongly constrain the shape of the lower end of the millisecond
pulsar period distribution, observations are needed that are not only
equally sensitive all pulsars slower than $\sim$0.5~ms, but that are
also sufficiently sensitive to detect a large number of pulsars.  The
work described here has shown that present technology is sufficient to
achieve the latter requirement through baseband processing technology,
however it falls short in basic sensitivity.  We expect that future
projects with cold, high-frequency receivers and larger
(Nyquist-sampled) bands will achieve the necessary sensitivity and
mitigation of scattering to finally resolve this issue.

\acknowledgements We thank S. Anderson for assistance tailoring CPSR
to our needs. RTE acknowledges the support of an Australian
Postgraduate Award. WvS was assisted by the Commonwealth Scholarship
and Fellowship program.  MB is an ARC Senior Research Fellow, and this
research was supported by the ARC Large Grants Scheme.


\clearpage

\begin{table}
\begin{center}
\caption{Observations of clusters with known pulsars}
\begin{tabular}{rllll}
\hline
\hline
Name & Refs & DM  & $N_{\rm obs}$ & $S_{\rm max}$\\
     &      & (pc~cm$^{-3}$) &    &  (mJy) \\
\hline
47 Tucanae & 1--3 & 23.4--25.4 & 5 & 3\\
M4 & 4 & 61.9--63.9 & 3 & \\
NGC 6342 & 5 & 74.9--76.9 & 2 & 1.3\\
NGC 6397 & 6 & 70.8--72.8 & 2 & \\
NGC 6440 & 7 & \ldots & 3 & \\
Terzan 5 & 8,9 & 237.0-242.1 & 3 & 5\\
NGC 6544 & 6 & 133.0--135.0 & 1& \\
NGC 6624 & 10 & 86.0--88.0 & 5& $9\pm 2$\\
M28      & 11 & 118.8--120.8 & 2& \\
NGC 6752 & 6 & 33.0--35.0 & 2& \\
\hline
\end{tabular}
\\
References: 
(1) \citealt{mld+90}
(2) \citealt{mlr+91};
(3) \citealt{clf+00};
(4) \citealt{lbb+88};
(5) \citealt{lbhb93};
(6) \citealt{dlm+00b};
(7) \citealt{mlj+89b};
(8) \citealt{lmd+90};
(9) \citealt{lmbm00};
(10) \citealt{bbl+94};
(11) \citealt{lbm+87}
\end{center}
\end{table}


\begin{table}
\begin{center}
\caption{Observations of clusters lacking pulsars}
\begin{tabular}{rl}
\hline
\hline
Name & $N_{\rm obs}$ \\
\hline
NGC 2808 & 3\\
E3 & 3 \\
NGC 3201 & 3\\
NGC 4372 & 4\\
NGC 4833 & 3\\
$\Omega$ Centaurus & 4\\
Liller 1 & 3\\
M22 & 2\\
M30 & 2\\
\end{tabular}
\end{center}
\end{table}

\begin{table}
\begin{center}
\caption{Detections}
\label{tab:det}
\begin{tabular}{llllll}
\hline
\hline
Name & Cluster & $P$ & DM & S/Ns \\
     &         & (ms)     & (pc~cm$^{-3}$) & \\
\hline
B0021--72D & 47 Tucanae & 5.35 & 24.7 & 34.3, 11.3\\
B0021--72E & 47 Tucanae & 3.54 & 24.2 & 15.4\\
B1620--26 & M4 & 11.1 & 62.9 & 23.7,22.8,20.5\\
B1744--24A & Terzan 5 &  11.6 & 242 &11.0 \\
B1820--30A & NGC 6624 &  5.44 & 86.8 &15.3 \\
B1821--24 & M28 & 3.05 & 120 & 10.4, 18.1\\
\hline
\end{tabular}
\end{center}
\end{table}


\end{document}